\documentclass[iop]{emulateapj}

\usepackage{color} 
\usepackage{ulem}
\usepackage{amsmath}
\usepackage{enumerate}
\usepackage{apjfonts}
\usepackage[colorlinks=true,citecolor=blue,breaklinks=true]{hyperref}
\usepackage[toc,page]{appendix}

\newcommand\cts{counts~s$^{-1}$}
\newcommand\ergs{erg~s$^{-1}$}
\newcommand\ergcms{erg~cm$^{-2}$~s$^{-1}$}

\begin{document}

\title{Super-Eddington accretion onto the Galactic ultraluminous X-ray pulsar Swift~J0243.6+6124}

\author{Lian Tao\altaffilmark{1}, Hua Feng\altaffilmark{2}, Shuangnan Zhang\altaffilmark{1,3,4}, Qingcui Bu\altaffilmark{1}, Shu Zhang\altaffilmark{1}, Jinlu Qu\altaffilmark{1}, Yue Zhang\altaffilmark{1,4}}

\altaffiltext{1}{Key Laboratory of Particle Astrophysics, Institute of High Energy Physics, Chinese Academy of Sciences, Beijing 100049, China}
\altaffiltext{2}{Department of Engineering Physics and Center for Astrophysics, Tsinghua University, Beijing 100084, China}
\altaffiltext{3}{National Astronomical Observatories, Chinese Academy of Sciences, Beijing 100012, China}
\altaffiltext{4}{University of the Chinese Academy of Sciences, Beijing, China}

\shorttitle{Swift~J0243.6+6124}
\shortauthors{Tao et al.}

\begin{abstract}

We report on the spectral behavior of the first Galactic ultraluminous X-ray pulsar Swift~J0243.6+6124 with \textit{NuSTAR} observations during its 2017-2018 outburst. At sub-Eddington levels, the source spectrum is characterized by three emission components, respectively from the accretion column, the hot spot, and a broad iron line emission region. When the source is above the Eddington limit, the hot spot temperature increases and the spectrum features two more blackbody components. One blackbody component has a radius of 10--20~km and is likely originated from the top of the accretion column. The other one saturates at a blackbody luminosity of $(1 - 2) \times 10^{38}$~\ergs, coincident with the Eddington limit of a neutron star. This is well consistent with the scenario that super-Eddington accretion onto compact objects will power optically-thick outflows and indicates an accretion rate 60--80 times the critical value. This  suggests that super-Eddington accretion onto magnetized systems can also power massive winds. At super-Eddington levels, the iron line becomes more significant and blueshifted, and is argued to be associated with the ultrafast wind in the central funnel or jets. This source, if located in external galaxies, will appear like other ultraluminous pulsars.

\end{abstract}


\section{Introduction}
\label{sec:intro}
The physical processes for super-Eddington accretion onto compact objects are still unclear. Ultraluminous X-ray sources (ULXs), which are non-nuclear point-like X-ray sources with apparent luminosities above $\sim$$10^{39}$~\ergs\ \citep[for a review see][]{Kaaret2017}, may be powered by super-Eddington accretion and are thus good targets for such studies  \citep[e.g.,][]{Middleton2015}. Among ULXs, the ultraluminous pulsars (ULPs) are of particular interest because the compact object mass is well constrained (1--3~$M_{\sun}$). So far, four ULPs have been detected, including M82 X-2 \citep{Bachetti2014}, NGC 7793 P13 \citep{Furst2016,Israel2017a}, NGC 5907 ULX1 \citep{Israel2017b} and NGC 300 ULX1 \citep{Carpano2018}, with a period of $\sim$1~s for the former three and $\sim$30~s for NGC 300 ULX1.  They all show a large spin-up rate, $\dot{\nu} \approx 10^{-10}$~Hz~s$^{-1}$, which is at least one order of magnitude greater than that of ordinary accreting pulsars, e.g., SMC X-1 \citep{Kahabka1999} and Cen X-3 \citep{van1980}.  The high spin-up rate of ULPs may be driven by their high accretion rates \citep{King2017}.  

Numerical simulations \citep{Ohsuga2011,Jiang2014,Sadowski2016} predict that super-Eddington accretion onto compact objects will launch nearly spherical, massive, optically-thick outflows/winds, as well as optically-thin ultra-fast outflows (UFOs) near the axis. Analytical analysis for super-Eddington accretion also reveals that massive outflows are inevitable \citep{Meier1982,Poutanen2007,Shen2015,Shen2016}. Blueshifted (0.1--0.2 $c$) absorption lines are seen in the high signal-to-noise X-ray spectra of ULXs \citep[e.g.,][]{Walton2016,Pinto2016,Pinto2017}, as direct evidence for the UFOs. The association of shock-ionized optical nebulae \citep{Pakull2003} with ULXs suggests the presence of the massive outflow interacting with the environment. The correlation between the line width and ionization level for broad emission lines in the optical spectra of ULXs is consistent with a wind origin \citep{Fabrika2015}. The soft excesses observed in the ULX spectra could be interpreted as due to emission from the photosphere of the optically-thick outflows \citep[e.g.,][]{Middleton2015,Weng2018}.  In addition to the uncollimated outflows, jets are also predicted by simulations under super-Eddington accretion \citep{Ohsuga2011,Narayan2017}.  SS~433 is believed to be accreting at a super-Eddington level and shows precessing and baryonic jets around 0.26~$c$ \citep{Fabrika2004}. The supersoft ULX M81 ULS-1 displays a varying (0.1--0.2~$c$) H{$\alpha$} emission line, which can be explained as due to jets similar to those seen in SS~433 \citep{Liu2015}. 

High quality spectroscopy is needed to test the models and simulation results, but is difficult for ULXs/ULPs due to their extragalactic distances. Swift J0243.6+6124 is a transient accreting pulsar in the Milky Way with a peak luminosity of $\sim 5 \times 10^{39}$~\ergs~\citep{Tsygankov2018}, exceeding the Eddington limit for a neutron star by a factor of $\sim$30, during its outburst in 2017-2018 \citep{Kennea2017}. The source exhibits a spin period of $\sim 9.86$~s \citep{Ge2017,Kennea2017,Jenke2017} and a spin-up rate of $\sim 10^{-10}$~Hz~s$^{-1}$ when $L_{\rm X} > 10^{39}$~\ergs\ \citep{Doroshenko2018}, which are similar to those of extragalactic ULPs. This offers us an opportunity to study a ULP at a Galactic distance.  Radio jets were observed during the outburst, which challenges the classical theory of jet formation with neutron stars \citep{van den Eijnden2018}, but is consistent with what expected for super-Eddington accretion as mentioned above. 

Here, we focus on the spectral modeling of Swift J0243.6+6124 with the \textit{NuSTAR} data, trying to test and constrain the physics for super-Eddington accretion. We adopt a distance of 7~kpc to the source measured with \textit{Gaia} \citep{Tsygankov2018,Wilson-Hodge2018}. 

\section{Observations and data reduction}
\label{sec:obs}

\tabletypesize{\scriptsize}
\begin{deluxetable}{lcll}
\tablecolumns{4}
\tablewidth{0pc}
\tablecaption{\textit{NuSTAR} Observations used in the paper \label{tab:obs}}
\tablehead{
\colhead{ObsID}  & \colhead{Date} & \colhead{Exposure} & \colhead{Count rate} \\
 \colhead{}        &   \colhead{}  & \colhead{(s)} & \colhead{(counts~s$^{-1}$)}
}
\startdata
90302319002 & 2017-10-05~15:51:09 & 14277 & $122.78\pm 0.09$ \\
90302319004 & 2017-10-31~07:21:09 & 1293  & $2700.3\pm 1.5$  \\
90302319006 & 2017-11-10~02:31:09 & 676  & $4001\pm 2$  \\
90302319008 & 2017-12-06~14:46:09 & 4589 & $959.5 \pm 0.5$ \\
90401308002 & 2018-03-10~12:21:09 & 27816 & $16.88 \pm 0.03$ 
\enddata
\tablecomments{Count rate is the net count rate of FPMA.}
\end{deluxetable}

\begin{figure}
\centering
\includegraphics[width=\columnwidth]{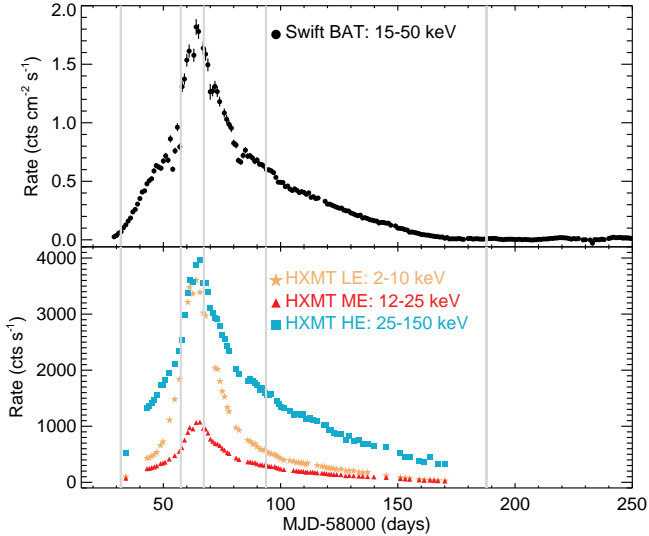} \\
\caption{X-ray lightcurves of Swift~J0243.6+6124 in the 2017--2018 outburst observed with \textit{Swift} and \textit{HXMT}. The five \textit{NuSTAR} observations are indicated by grey vertical lines.   
\label{fig:cts}}
\end{figure}

We collected five \textit{NuSTAR} \citep{Harrison2013} observations during the 2017-2018 outburst of Swift J0243.6+6124 (Table~\ref{tab:obs}). A very short observation (ObsID 90401308001) with an exposure time of 18~s was discarded.  The times of the \textit{NuSTAR} observations were marked in Figure~\ref{fig:cts}, on top of the \textit{Swift}/BAT\footnote{Data obtained from https://swift.gsfc.nasa.gov/results/\\transients/weak/SwiftJ0243.6p6124/} and \textit{Insight}-HXMT\footnote{Data adopted from Zhang et al.\ (2019), in preparation} \citep{Zhang2014} lightcurves. The five \textit{NuSTAR} observations sampled the complete cycle of the outburst at different epochs, composing a treasure database for the study of ULPs.

The cleaned \textit{NuSTAR} event files are created using NuSTARDAS pipeline 1.8.0 in HEASoft v6.22, with the \textit{NuSTAR} CALDB version 20180419. The peak flux of Swift J0243.6+6124 is up to $\sim 8$~Crab\footnote{https://swift.gsfc.nasa.gov/results/transients/}, in which case some source events may be improperly discarded by a noise filter in {\tt nupipeline}, resulting in an underestimate of the source flux.\footnote{https://heasarc.gsfc.nasa.gov/docs/nustar/analysis/} Therefore, following the analysis guide, we first extracted the light curves in 1-second bins, and altered the filter in the four observations (except ObsID 90401308002) where the count rate exceeds 100~\cts. This was done by setting the keyword {\tt statusexpr} with an expression of ``STATUS==b0000xxx00xxxx000'' before the cleaned event files were generated. Then, the source events were extracted from a circular region centered at the source position with a radius of 180\arcsec. The background was estimated from a nearby source-free circular region. Energy spectra were binned to have at least 50 counts per bin.  

\section{Spectral modeling and results}
\label{sec:res}

\begin{deluxetable*}{lllllll}
\tablecolumns{7}
\tablewidth{0pc}
\tablecaption{Best-fit models and spectral parameters \label{tab:fit}}
\tablehead{
 \colhead{Component} & \colhead{Parameter} & \colhead{90302319002} & \colhead{90302319004} & \colhead{90302319006} & \colhead{90302319008} & \colhead{90401308002} 
}
\startdata
{\tt tbabs} & $N_{\rm H}~(10^{22}~\rm cm^{-2})$ & $1.0\pm 0.2$  &  $0.6^{+2.8}_{-0.6}$ & $6^{+2}_{-5}$ & $ 2.4^{+1.0}_{-0.9}$ & $1.7 \pm 0.5$ \\
{\tt cutoffpl} (column) & $\rm \Gamma$ &  $1.11\pm 0.03$  &  $1.1^{+0.3}_{-0.4}$ & $1.4^{+0.3}_{-0.5}$ & $0.42^{+0.15}_{-0.20}$ & $1.21 \pm 0.05$  \\
  & $E_{\rm cut}$ (keV) & $24.5\pm 0.9$  &  $20^{+3}_{-2}$ & $26^{+8}_{-5}$ & $17.4\pm 1.3$  & $25.2^{+1.8}_{-1.6}$ \\
  & $N_{\rm cut}$ (photons~keV$^{-1}$~cm$^{-2}$~s$^{-1}$)  & $0.282^{+0.013}_{-0.012}$ &  $4\pm 3$ & $14^{+15}_{-7}$ & $0.40^{+0.20}_{-0.17}$  & $0.052^{+0.006}_{-0.005}$ \\ 
{\tt bbodyrad} (hot spot) & $T_{\rm bb}$ (keV) & $3.06\pm 0.03$  &  $4.5\pm 0.4$ & $4.4^{+0.5}_{-0.3}$ & $4.43^{+0.18}_{-0.16}$  & $2.21 \pm 0.06$ \\
  & $R_{\rm bb}$ (km) & $0.997^{+0.015}_{-0.014}$  & $1.4^{+0.7}_{-0.3}$ & $1.8^{+0.8}_{-0.3}$ & $1.28^{+0.18}_{-0.13} $ & $0.57^{+0.04}_{-0.03}$ \\
{\tt bbodyrad} (column top) & $T_{\rm bb}$ (keV) & \nodata  &  $1.46\pm 0.07$ & $1.40^{+0.07}_{-0.04}$ & $1.435^{+0.018}_{-0.017}$ & \nodata \\
  & $R_{\rm bb}$ (km) & \nodata &  $20\pm 3$ & $27 \pm 4$ & $12.5 \pm 0.4$ & \nodata \\
{\tt bbodyrad} (outflow) & $T_{\rm bb}$ (keV) &  \nodata  &   $0.57^{+0.11}_{-0.09}$ & $0.42^{+0.09}_{-0.03}$ & $0.481\pm 0.014$ & \nodata \\
  & $R_{\rm bb}$ (km) & \nodata &  $120^{+170}_{-30}$  & $500^{+300}_{-200}$  &  $109^{+16}_{-14}$ & \nodata \\
  & $L_{\rm bb}~(10^{38}$~\ergs) &  \nodata  & $1.88^{+1.96}_{-0.15}$ & $8 \pm 6 $ & $ 0.82^{+0.19}_{-0.17}$  &  \nodata \\
{\tt gaussian} (iron) & $E_{\rm g}$ (keV) & $6.44\pm 0.04$ &  $6.80^{+0.16}_{-0.20}$ & $6.99^{+0.12}_{-0.19}$ & $6.476 \pm 0.017$ & $6.33^{+0.05}_{-0.09}$ \\
  & $\sigma$ (keV) & $0.45^{+0.07}_{-0.06} $ & $1.26^{+0.13}_{-0.11} $ & $1.14^{+0.14}_{-0.10}$ & $0.27 \pm 0.03$  & $0.34^{+0.17}_{-0.14}$ \\
  & $EW$ (keV) &  $0.080^{+0.008}_{-0.009}$ & $ 0.51^{+0.51}_{-0.19} $ & $ 0.36^{+0.16}_{-0.10} $ & $0.058^{+0.009}_{-0.007}$ & $0.076 \pm 0.012$ \\
{\tt edge} (iron) & $E_{\rm edge}$ (keV) & \nodata  & $7.027^{+0.018}_{-0.025}$ & $6.99 \pm 0.05$ & \nodata  & \nodata \\
  & $\tau$ &  \nodata & $0.156^{+0.013}_{-0.021} $ & $ 0.097^{+0.030}_{-0.019}  $ & \nodata & \nodata \\
\noalign{\smallskip}\hline\noalign{\smallskip}
 & $F_{\rm 3-79 keV}$ (\ergcms) & $8.73\times 10^{-9}$  &  $ 1.55 \times 10^{-7}$  & $ 2.27 \times 10^{-7}$  & $ 0.74 \times 10^{-7}$ & $1.14\times 10^{-9}$ \\
 & $L_{\rm 0.1-100 keV} $ (\ergs) & $6.01 \times 10^{37}$ & $ 1.22 \times 10^{39}$  &  $2.67 \times 10^{39}$ & $ 0.55 \times 10^{39}$ & $ 0.83 \times 10^{37}$ \\
\noalign{\smallskip}\hline\noalign{\smallskip}
  & $\chi^2/{\rm dof}$ & 2084.7/1987 &  2207.1/1986 & 2040.5/1909 & 2811.3/2479 & 1476.9/1476
\enddata
\tablecomments{$N_{\rm cut}$ is the {\tt cutoffpl} normalization at 1 keV;
$F_{\rm 3-79 keV}$ is the observed flux of FPMA and $L_{\rm 0.1-100 keV}$ is the absorption corrected luminosity. All errors are quoted at 90\% confidence level.}
\end{deluxetable*}

\begin{figure}
\centering
\includegraphics[width=\columnwidth]{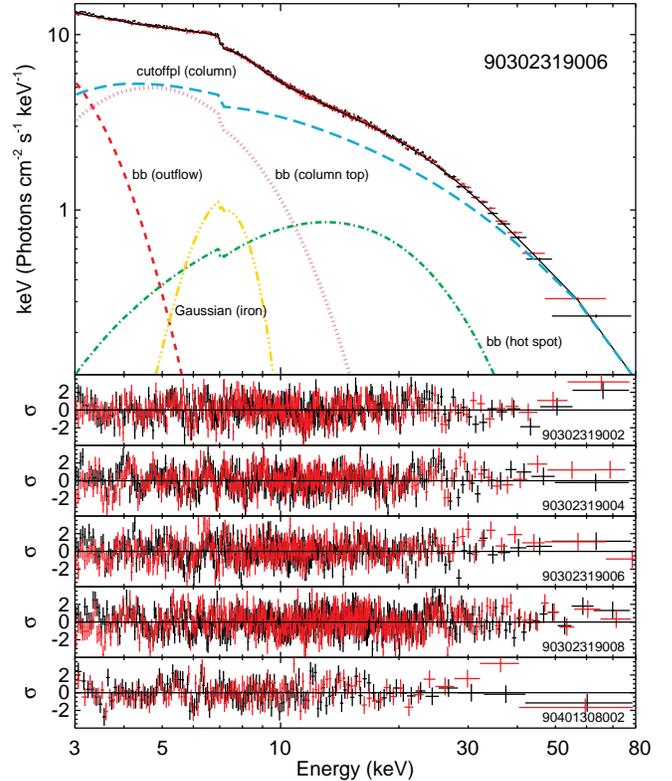} 
\caption{Top: energy spectra and model components. Bottoms: spectral residuals with respect to the best-fit model. 
\label{fig:model}}
\end{figure}

The spectral fit was done in XSPEC v12.10.0e \citep{Arnaud1996}. Following \citet{Bahramian2017} and \citet{Jaisawal2018}, we first adopted a three-component model ({\tt model 1 =  cutoffpl + bbodyrad + gaussian}) subject to interstellar absorption ({\tt tbabs}). The three components are to account for emission from the accretion column, the thermal emission from the hot spot around the polar region, and the iron emission region, respectively.  {\tt Model 1} gives an adequate fit to the two spectra in the low states, from the first and last observations, with model parameters consistent with those reported in \citet{Jaisawal2018} for ObsID 90302319002. For the remaining three observations, which were taken around the peak of the outburst when the luminosity was above the Eddington limit, {\tt model 1} is unable to fit the spectra successfully and a more complicated model is needed. Following the physical picture for super-Eddington accretion, where optically-thick outflows may be launched \citep{Zhou2019} and the accretion column may be extended \citep{Mushtukov2015}, we added two additional thermal components ({\tt model 2 = model 1 + bbodyrad + bbodyrad}), respectively, to account for the thermal emission from the photosphere of the outflow and from the extended column. Moreover, for the two observations (ObsID 90302319004 and 90302319006) with the highest luminosities, an additional {\tt edge} component is required near the iron K edge. The spectral parameters are listed in Table~\ref{tab:fit} and the model components are plotted in Figure~\ref{fig:model}.  

\begin{figure}
\centering
\includegraphics[width=0.8\columnwidth]{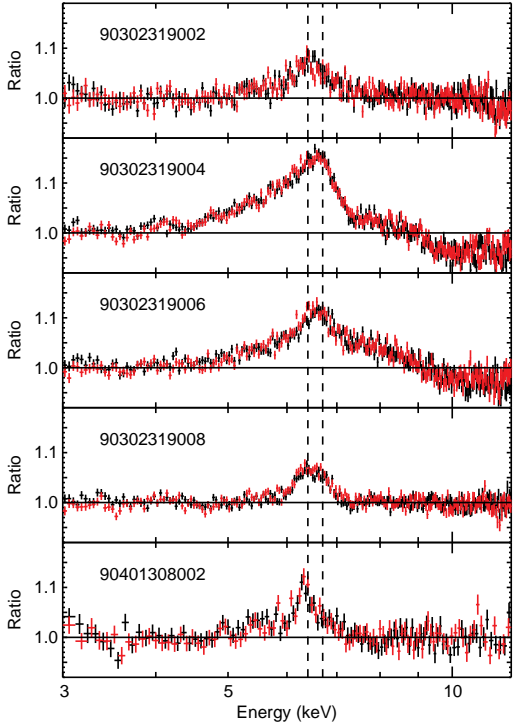} 
\caption{Data to model ratios with the Gaussian and absorption edge removed. The vertical lines mark 6.4~keV and 6.7~keV, respectively, at the K$\alpha$ energies of Fe~I and Fe~XXV.
\label{fig:iron}}
\end{figure}

A minor feature in the residuals near 6-7 keV for ObsIDs 90302319004 and 90302319006 may be seen. It is likely related to the asymmetry of the iron line, but is weak and will not influence our results here. In order to inspect the broad iron line, we removed the emission line and absorption edge components from the model, and plotted the residuals in Figure~\ref{fig:iron} to enhance the iron line feature.

\section{Timing analysis}
\label{sec:timing}

\begin{figure}
\centering
\includegraphics[width=\columnwidth]{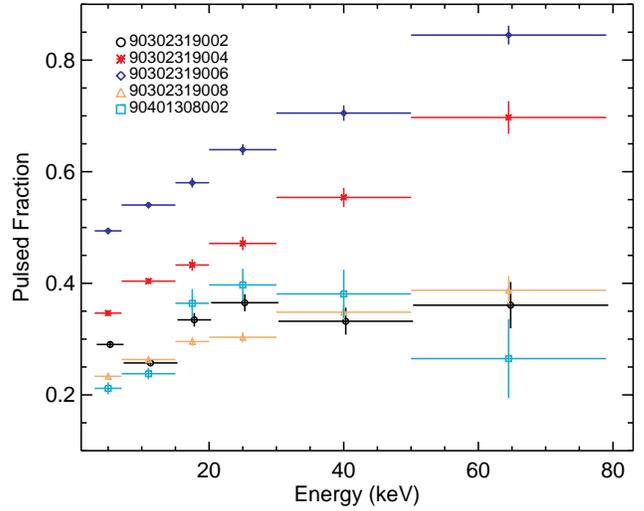} 
\caption{Pulsed fraction versus X-ray energy. 
\label{fig:fraction}}
\end{figure}

The barycenter correction was applied before the light curves were binned to a time resolution of 100~ms. We searched the spin period using the {\tt efsearch} task in {\tt FTOOLS}, and found that the spin period decreased from 9.8539~s to 9.7933~s from the first to the last observation. For each observation, the pulse profile was obtained by folding the light curves with the {\tt FTOOLS} {\tt efold} task. The pulsed fraction are calculated in different energy bands (Figure~\ref{fig:fraction}), defined as $(F_{\rm max}-F_{\rm min})/(F_{\rm max}+F_{\rm min})$, where $F_{\rm max}$ and $F_{\rm min}$ are maximum and minimum intensity, respectively. For the three observations at a super-Eddington level, the pulsed fraction increases with the increasing photon energy and accretion rate. For the remaining two observations, it peaks around 20 keV. The results in the low enengy bands are consistent with those obtained with {\it NICER} data \citep{Wilson-Hodge2018}.

\section{Discussion}
\label{sec:dis}

\begin{figure}
\includegraphics[width=0.99\columnwidth,angle=270]{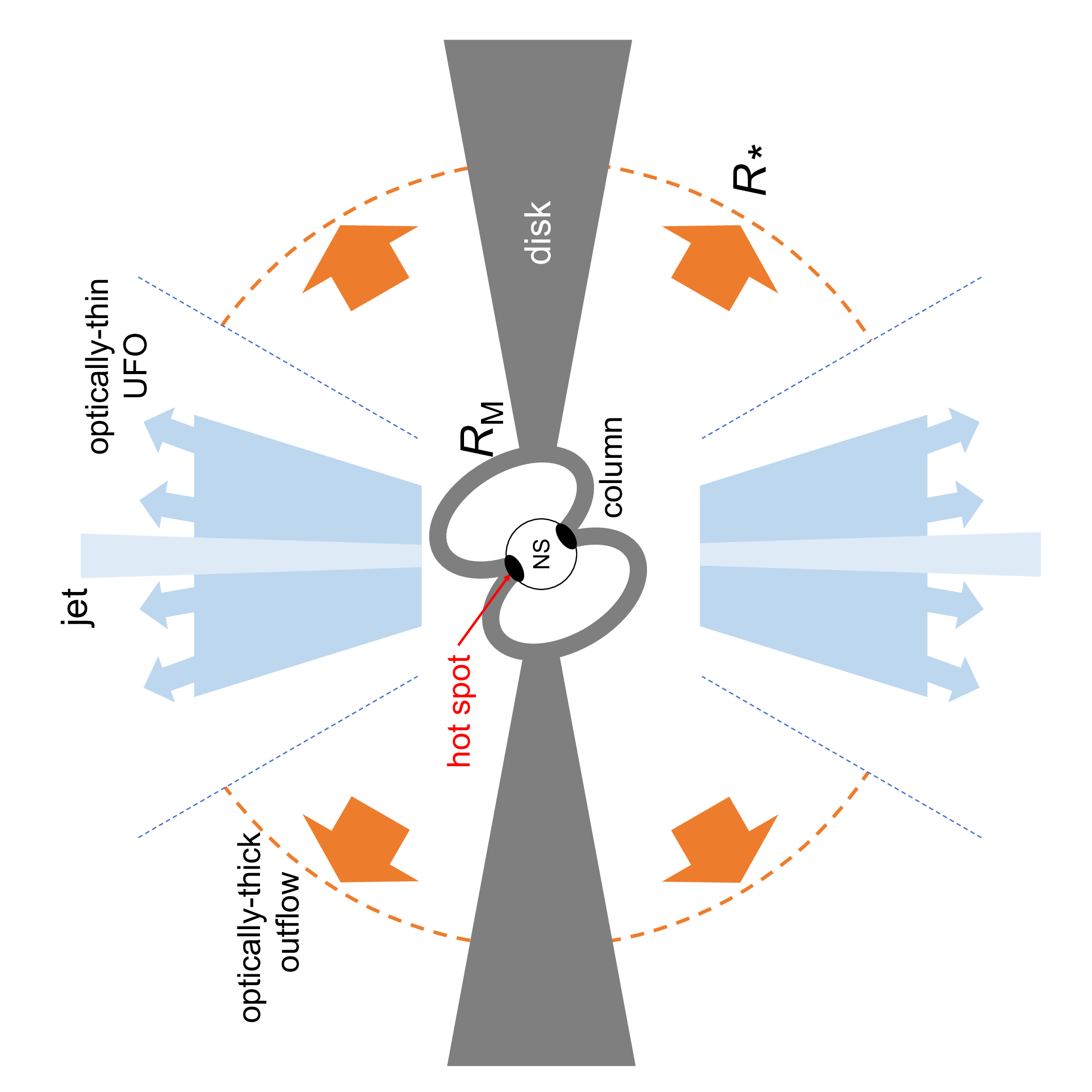} 
\caption{Schematic drawing of the physical picture for super-Eddington accretion (not to scale). 
\label{fig:schematic}}
\end{figure}

Swift J0243.6+6124 is the first known ULP in our Galaxy, enabling a close look at the properties of the same kind.  In the following, we will interpret the spectral results in a scenario consistent with the super-Eddington accretion. A schematic drawing of the physical picture can be found in Figure~\ref{fig:schematic}. 

When the source accretes at a sub-Eddington level, only three emission components were detected, respectively from the accretion column, the iron line, and the hot spot (blackbody, $T_{\rm bb} = 2-3$~keV, $R_{\rm bb} = 0.6-1.0$~km). When the luminosity is in excess of the Eddington limit, two additional thermal components, sometimes along with an absorption edge, are needed to fit the spectra. In these cases, the hot spots become even hotter, with a temperature of about 4.5~keV but the radius remains around 1~km. One of the new thermal components has a temperature of around 1.5~keV and a radius of 10--20~km. This could be explained by emission from the top of the accretion column at super-Eddington level, where the accretion column may grow to a size comparable to the neutron star radius, with a temperature gradient, from 4.5~keV at the base to 1.5~keV at the top \citep{Mushtukov2015}. A two-temperature blackbody is perhaps an approximation of the thermal emission from the base to the top.

The other new thermal component has a temperature of about 0.5--0.6 keV with a radius of 110--120~km, or a blackbody luminosity of about (1--2)$\times 10^{38}$~\ergs. The blackbody luminosity stays constant between observations. We note that the blackbody luminosity derived from ObsID 90302319006 is consistent with this value but cannot be well constrained. The luminosity of this thermal component is coincident with the Eddington limit of a neutron star. This can be self-consistently explained if the thermal emission arises from the photosphere of the optically-thick outflow driven by super-Eddington accretion. According to the super-Eddington accretion models, the outflow is launched near the radius where the Eddington limit is approached \citep{Shakura1973,Meier1982,Poutanen2007,Shen2015,Shen2016}. If the wind launch radius is above the photon trapping radius, the radiation and materials in the wind will stay in local thermodynamic equilibrium and the radiative luminosity will stay constant at the Eddington limit. Thus, the massive, optically-thick wind under super-Eddington accretion is thought to be Eddington limited, which was found to be consistent with observations for supersoft ULXs \citep{Urquhart2016,Feng2016,Zhou2019}. Thus, the thermal component detected here can be naturally explained as a signature of the optically-thick outflow. 

Following the outflow model of \citet{Meier1982} \citep[see more details in][]{Zhou2019}, assuming a $1.4~M_\odot$ neutron star, we found that one needs a mass accretion rate of $\dot{m} = (60-80)~\dot{M} / \dot{M}_{\rm Edd}$ to match the observed blackbody temperature ($T_{\rm bb} \approx 0.5-0.6$~keV), where $\dot{M}_{\rm Edd} = L_{\rm Edd} / 0.1c^2 $ is the critical mass accretion rate needed to power the Eddington limit.  The model predicts a scattering optical depth ($\tau_{\rm es}$) at the photosphere of around 200. In this case,  the physical radius of the photosphere is $R_\ast \approx R_{\rm bb} \sqrt{\tau_{\rm es}} \approx 1.6 \times 10^3 $~km. The model assumes that the wind is launched at a radius ($R_{\rm i}$) near the advective radius of a slim disk, $R_{\rm i} \approx 750 - 1000 \; {\rm km}$.  The accretion disk is truncated by the magnetic fields at $R_{\rm M} = (3.3 \times 10^2 \; {\rm km}) \; B_{12}^{4/7} L_{39}^{-2/7} R_6^{10/7}  m_{1.4}^{1/7}$ \citep{Ghosh1979}. If the magnetic field of Swift J0243.6+6124 is $10^{13}$~G \citep{Doroshenko2018,Wilson-Hodge2018}, one gets $R_{\rm M} \approx 1 \times 10^{3}$~km with typical values for other parameters; if we follow \citet{Tsygankov2018} and use the upper limit for the magnetic fields ($B < 6.2 \times 10^{12}$~G) derived from the undetected propeller effect, one gets $R_{\rm M} < 900$~km. We note that these estimates are precise only to the order-of-magnitude, but it suggests that the optically-thick wind may have been launched at a position close to the Alfv{\'e}n radius. 

The broad iron emission line is significant in the spectra of Swift J0243.6+6124. Similar features have been frequently observed in low mass neutron star X-ray binaries, such as Serpens X-1 \citep{Bhattacharyya2007}, 4U 1820-30, GX 349+2 \citep{Cackett2008}, and GX 17+2 \citep{Ludlam2017}. None of the extragalactic ULPs shows such a feature.  We ran simulations and found that this is due to observational effect. If one takes the spectral model of Swift J0243.6+6124 in the outburst peak, where the iron line is most prominent, renormalizes the X-ray luminosity to $10^{40}$ \ergs\ in 0.3-10 keV, and places the source at a distance of 1.9~Mpc as for the closest ULP (NGC 300 ULX), the iron line is not detectable with an \textit{XMM-Newton} observation even with an exposure of 100~ks. 

As one can see from Figure~\ref{fig:iron}, the line strength is correlated with the X-ray luminosity, suggesting that the line may be generated in association with the winds or jets driven by super-Eddington accretion. In the first observation, the peak energy is consistent with 6.4~keV expected for the K$\alpha$ emission from Fe~I; in the second and third observations during the outburst peak, it is higher than 6.4~keV but less than 6.7~keV expected for Fe~XXV, which is detected in the jets of SS~433 \citep{Kotani1994,Marshall2002}; in the fourth observation, it seems to have double peaks or a flat top, but the evidence is marginal; in the last observation, a narrow component below 6.4 keV seems to stand out above a broad component. The Fe absorption edge appears only in the most luminous states, implying that they are associated with the optically-thin part of the wind, perhaps the UFOs in the funnel. 

One possibility is that, the materials that have iron emission is mostly neutral at relatively low luminosities, but the ionization level increases towards high luminosities. However, this cannot explain the fact that the narrow component in the last observation is below the neutral iron K$\alpha$ energy. A more likely interpretation is that, the narrow iron emission originates in the jets. The red-shifted narrow line in the last observation arises from the receding jet, while the possibly double-peaked feature in the fourth observation is due to both approaching and receding jets. In the super-Eddington regime, the UFO in the central funnel may have a speed close to the jet (0.1--0.2~$c$) and produce a velocity broadened, blue-shifted iron line, in line with what we observed in the second and third observation. These speculations may be tested with future high-resolution spectroscopic observations.

In our interpretation, the emission from the accretion column will cause X-ray pulsations, mainly from the cutoff power-law component. The thermal emission from the hot spot and the column is also modulated but perhaps at a lower degree. Emission from the outflow should not be modulated by the rotation of the neutron star. As the cutoff power-law component dominates towards high energies, the pulsed fraction is expected to be stronger at higher energies, which is in good agreement with measurements during the super-Eddington level. However, when the source is at sub-Eddington, why the pulsed fraction drops above 20~keV is puzzling, which may suggest a geometry different from super-Eddington accretion. In this paper, we mainly focus on the spectral properties. How to link the spectral behavior to the timing properties quantitatively needs in-depth investigations.

\begin{deluxetable}{llll}
\tablecolumns{4}
\tablewidth{0pc}
\tablecaption{Spectral fitting if the source is moved to an extragalactic distance of 1.9~Mpc.\label{tab:comparison}} 
\tablehead{
 \colhead{Component} & \colhead{Parameter} & \colhead{90302319004} & \colhead{90302319006} 
}
\startdata
{\tt tbabs} & $N_{\rm H}~(10^{22}~\rm cm^{-2})$ & $4.3^{+1.0}_{-0.9}$  &  $6.0\pm 1.3$  \\
{\tt highecut} & $E_{\rm c}$ (keV) & $6.6^{+0.7}_{-0.6}$  &  $6.4^{+0.8}_{-0.6}$  \\
  & $E_{\rm f}$ (keV) & $63^{+29}_{-14}$  &  $43^{+19}_{-10}$  \\
{\tt powerlaw} & $\rm \Gamma$ &  $2.10^{+0.07}_{-0.06}$  &  $2.01\pm 0.10$   \\
\noalign{\smallskip}\hline\noalign{\smallskip}
  & $\chi^2/{\rm dof}$ & 654.0/531 &  412.8/431 \\
\noalign{\smallskip}\hline\noalign{\smallskip}
{\tt tbabs} & $N_{\rm H}~(10^{22}~\rm cm^{-2})^{a}$ & $5$  &  $5$  \\
{\tt diskbb} & $T_{\rm in}$ (keV) & $0.47\pm 0.05$  &  $0.37^{+0.09}_{-0.07}$  \\
  & $R_{\rm in}$ (km)$^{b}$ & $460^{+360}_{-130}$  & $1300^{+6000}_{-500}$  \\
{\tt bbodyrad} & $T_{\rm bb}$ (keV) & $1.37\pm 0.04$  &  $1.35\pm 0.05$  \\
  & $R_{\rm bb}$ (km) & $28.1^{+1.9}_{-1.8}$  & $34^{+3}_{-2}$ \\
{\tt cutoffpl} & $E_{\rm cut}$ (keV) & $12.6^{+0.7}_{-0.6}$  &  $11.8^{+0.8}_{-0.7}$  \\
\noalign{\smallskip}\hline\noalign{\smallskip}
  & $\chi^2/{\rm dof}$ & 555.9/530 &  353.8/430 
\enddata
\tablecomments{$^{a}$$N_{\rm H}$ is hard to be constrained and fixed at $5 \times 10^{22}~\rm cm^{-2}$. $^{b}$Following \citet{Walton2018}, $R_{\rm in}$ is obtained assuming cos~$\theta=f_{\rm col} =1$. All errors are quoted at 90\% confidence level. }
\end{deluxetable}

An interesting question is whether or not this source appears like extragalactic ULPs. When Swift J0243.6+6124 is in the ultraluminous phase (in the second and third observations), we extract a short-exposure spectrum by randomly selecting part of the photons as if the source has a luminosity of $10^{40}$ \ergs\ in 0.1--100 keV, observed with an exposure of 100~ks at a distance of 1.9~Mpc (distance to the nearest extragalactic ULP in NGC 300). The spectra were then fit with two models that have been used for ULPs,  power-law with high energy cutoff \citep{Pintore2017} and a three-component model, cool disk blackbody +  hot blackbody + cutoff power-law \citep{Walton2018}. The blackbody component used in \citet{Pintore2017} has a temperature too low to be detectable with {\it NuSTAR}. To be consistent with \citet{Walton2018}, we froze the power-law photon index at 0.5. Both models provide a good fit, although the first model results in some residuals around 5--10~keV, see Table~\ref{tab:comparison} for best-fit parameters. The derived parameters are consistent with those obtained for extragalactic ULPs, except that the e-folding energy is higher for Swift J0243.6+6124, suggestive of a higher temperature. In the second model, the high temperature blackbody could arise from the accretion column and the cool thermal component (though fitted with a disk blackbody) is likely from the outflow. We note that the hot spot and the iron line component cannot be significantly detected in the spectra.

To summarize, Swift J0243.6+6124 has offered us a unique opportunity to have a close look at the super-Eddington accretion onto a neutron star. In this paper, we proposed a physical picture that can self-consistently explain the spectral behavior of the source, suggesting that the massive, optically-thick outflows can be launched in magnetized systems under super-Eddington accretion. 

\acknowledgements 

We thank the anonymous referee, Diego Altamirano, and Can G\"{U}NG\"{O}R for useful comments. LT acknowledges funding support from the National Natural Science Foundation of China (NSFC) under grant numbers U1838115 and U1838201, the CAS Pioneer Hundred Talent Program Y8291130K2 and the Scientific and technological innovation project of IHEP Y7515570U1. HF acknowledges funding support from the National Key R\&D Program of China (grant Nos.\ 2016YFA040080X and 2018YFA0404502), and the National Natural Science Foundation of China under the grant Nos.\ 11633003 and 11821303. SZ thanks support from the Chinese NSFC U1838201, 11733009 and 11473027. JQ thanks support from the Chinese NSFC 11673023.

{\it Facilities:} \textit{NuSTAR}, \textit{Insight}-HXMT, \textit{Swift}

\clearpage

\end{document}